\documentstyle[preprint,aps,epsfig]{revtex}
\tighten
\newcommand{\beq}{\begin{eqnarray}}
\newcommand{\eeq}{\end{eqnarray}}

\newcommand{\bpp}{B_d^0 \to \pi^+ \pi^-}
\newcommand{\spp}{S_{\pi\pi} }
\newcommand{\cpp}{C_{\pi\pi} }
\newcommand{\dpp}{D_{\pi\pi} }

\newcommand{\lpp}{\lambda_{\pi\pi} }
\newcommand{\app}{A_{\pi\pi} }
\newcommand{\ppp}{P_{\pi\pi} }
\newcommand{\tpp}{T_{\pi\pi} }
\newcommand{\rhob}{\bar{\rho} }
\newcommand{\etab}{\bar{\eta} }

\newcommand{\ssb}{\sin{(2 \beta)} }

\newcommand{\im}{{\rm Im}\,}
\newcommand{\non}{\nonumber\\ }
\def \epjc{  Eur. Phys. J. C }

\def \npb{  Nucl. Phys. B }
\def \plb{  Phys. Lett. B }
\def \prd{  Phys. Rev. D }
\def \prl{  Phys. Rev. Lett.  }

\begin{document}
\preprint{BIHEP-TH-2002-22}

\title{Constraint on the CKM angle $\alpha$ from the experimental measurements
of CP violation in $B_d^0 \to \pi^+ \pi^-$ decay }
\author{
 Cai-Dian L\"u$^1$\thanks{Email address: lucd@ihep.ac.cn}
 and  Zhenjun Xiao$^2$\thanks{Email address: zjxiao@email.njnu.edu.cn}
 \\
{\small $^1$ CCAST (World Laboratory), P.O. Box 8730, Beijing 100080, China;}\\
{\small $^1$ Institute of High Energy Physics,  P.O. Box 918(4),
Beijing 100039, China
\thanks{Mailing address.}}    \\
{\small $^2$ Department of Physics, Nanjing Normal University,
Nanjing, Jiangsu 210097, China}                          }
\date{\today}
\maketitle
\begin{abstract}
In this paper, we study and try to find the constraint on the CKM
angle $\alpha$ from the experimental measurements of CP violation
in $B_d^0 \to \pi^+ \pi^-$ decay, as reported very recently by
BaBar and Belle Collaborations. After considering uncertainties of
the data and the ratio  $r$ of penguin over tree amplitude, we
found that strong constraint on both the CKM angle $\alpha$ and
the strong phase $\delta$ can be obtained from the measured CP
asymmetries $\spp$ and $\app$: (a) the ranges of $87^\circ \leq
\alpha \leq 131^\circ$ and $36^\circ \leq \delta \leq 144^\circ$
are allowed by $1\sigma$ of the averaged data for $r =0.31$; (b)
for Belle's result alone,  the limits on $\alpha$ and $\delta$ are
$ 104^\circ \leq \alpha \leq  139^\circ$ and $42^\circ \leq \delta
\leq 138^\circ$ for $ 0.32 \leq r \leq 0.41$; and (c) the
angle $\alpha$ larger than $90^\circ$ is preferred.
\end{abstract}


\newpage
\section{Introduction}

 To study CP violation mechanism is one of the main goals of the B
 factory experiments.
In the standard model (SM), the CP violation is induced by the nonzero phase angle appeared in
the   Cabbibo-Kobayashi-Maskawa  (CKM) mixing matrix.  Recent measurements of $\sin{2\beta}$ in neutral B meson decay
$B_d^0 \to J/\psi K_{S,L}^0$ by BABAR \cite{87-091801,babar-0203007} and Belle
\cite{87-091802,belle-0202027} Collaborations
established the third type CP violation (interference between the decay and mixing ) of
$B_d$ meson system. Two new measurements of $\sin{2\beta}$ as reported this year
by BaBar \cite{babar-0203007} and Belle \cite{belle-0204002} Collaborations are
\beq
\ssb &=& 0.75 \pm 0.09(stat.) \pm 0.04 (syst.), \label{eq:ssb-babar}  \\
\ssb &=& 0.82 \pm 0.12(stat.) \pm 0.05 (syst.), \label{eq:ssb-belle}
\eeq
with an average
\beq
\ssb = 0.78 \pm 0.08, \label{eq:ssb-av2}
\eeq
which is well consistent with last year's world average, $\ssb = 0.79 \pm 0.10$,
and leads to the bounds on the angle $\beta$:
\beq
\beta = \left ( 26^\circ \pm 4^\circ\right )  \bigvee
\left (  64^\circ \pm 4^\circ \right ). \label{eq:b-beta}
\eeq

Despite the well measured CKM angle $\beta$, we have very poor
knowledge on the other two angles $\alpha$ and $\gamma$.
Very recently, Belle collaboration reported their first
measurements of the CP violation of the $\bpp$ decay
\cite{belle-0204002}:
\beq \spp &=& -1.21 ^{+0.38}_{-0.27} (stat.)^{+0.16}_{-0.13}(syst.), \non
A_{\pi\pi} &=& +0.94^{+0.25}_{-0.31} (stat.) \pm 0.09 (syst).
\label{eq:belle}
\eeq
The probability for
$\spp \neq 0$ and $A_{\pi\pi} \neq 0$ is $99.9\%$ \cite{belle-0204002}.
Based on a data sample of about $88$ million $\Upsilon(4S) \to B\bar{B}$ decays,
the BaBar Collaboration updated their measurement of CP violating asymmetries
of $B \to \pi^+ \pi^-$ decay
\footnote{For the parameter $A_{\pi\pi}$, $C_{\pi\pi}$, there is a sign difference
between the conventions of Belle and BaBar Collaboration  $A_{\pi\pi}=-C_{\pi\pi}$.
We here use  Belle's convention\cite{belle-0204002}. }
\cite{babar-0207055}
\beq
\spp &=& 0.02 \pm 0.34 (stat.) \pm 0.05(syst.), \non
C_{\pi\pi} &=& -0.30 \pm 0.25 (stat.) \pm 0.04(syst.).\label{eq:babar}
\eeq
The uncertainties of BABAR's new results are smaller than those of their
previous results\cite{65-051502,babar-0205082}. It is easy to see that the
experimental measurements of BABAR
and Belle collaborations are not fully consistent with each other: BABAR's results are still
consistent with zero, while the Bells's results strongly indicate nonzero $\spp$ and $\app$.
Further improvement of the data will enable us to draw definite conclusions about the values of
both $\spp$ and $\app$.

Inspired by the recent measurements, discussions have been made to
obtain information on strong phases and CKM phases from the
recent experimental measurements \cite{rosner,0204101}.
In Ref.\cite{rosner}, Gronau and Rosner examined the time-dependent measurements of
$B \to \pi^+ \pi^-$ decay to draw information on strong and weak phases and
found that: (a) if $\sin{\delta}$ is small a discrete ambiguity between $\delta \simeq 0$
and $\delta \simeq \pi$ could be resolved by comparing the measured branching
ratio $Br(B \to \pi^+ \pi^-)$ with that predicted in the absence of the
penguin amplitude;
(b) if $A_{\pi\pi}$ is non-zero, the discrete ambiguity between $\delta$ and
$\pi -\delta$ becomes harder to resolve, but its effects on CKM parameters
becomes less important,
and (c) the sign of the quantity $\dpp=2Re(\lambda_{\pi\pi})/(
1 + |\lambda_{\pi\pi}|^2)$ is always negative for the allowed range of CKM
parameters and therefore a positive value of $\dpp$ would signify new physics beyond the
SM. In Ref.\cite{0204101}, Fleischer and Matias investigated the allowed regions in observable
space of $B\to \pi K$, $B_d \to \pi^+ \pi^-$ and $B_s \to K^+ K^-$ decays. They
considered the correlations between these three kinds of decay modes implied by the
SU(3) flavor symmetry and
the U-spin symmetry and found the new constraint on the CKM angle $\gamma$ by using the B-factory
measurements of CP violation of $B_d \to \pi^+ \pi^-$.

As it is well-known that, the CP asymmetry measurements in $B\to \pi \pi$ decays play an important role
in extracting out the CKM angle $\alpha$. In this paper, we focus on the $B \to \pi^+ \pi^-$
decay and try to extract out the constraint on the angle $\alpha$ from  the measured $\spp$
and $A_{\pi\pi}$ and the ratio $r$ of penguin over tree amplitude fixed by theoretical argument.
Taking into account both Belle and BABAR newest measurements\cite{belle-0204002,babar-0207055},
the weighted-averages of $\spp$ and $A_{\pi\pi}$ are
\beq
\spp^{exp} = -0.57  \pm 0.25, \ \
\app^{exp} = 0.57 \pm 0.19 . \label{eq:data}
\eeq
We will treat above averages as the measured asymmetries of $B_d \to \pi^+ \pi^-$ decay
in the following analysis. We also investigate what will happen if only  Bells's measurements
are taken into account. For
the case of $\spp\approx 0$ and $\app \approx 0$ as indicated by BABAR's results
alone, one can see the discussions given in Ref.\cite{rosner}.

This paper is organized as follows. In Sec.~\ref{sec-2} we present the general
description of CP asymmetries of the $B \to \pi^+ \pi^-$ decay.
In Sec.~\ref{sec-3} we consider the new Babar and Belle's measurements of $\spp$ and $\app$
to draw the constraints on the CKM angle $\alpha$ and the strong phase $\delta$.
The conclusions are included in the final section.

\section{CP asymmetries of $B \to \pi^+ \pi^-$ decay} \label{sec-2}

In the SM with SU(2)$\times$U(1) as the gauge group, the quark mass eigenstates are not the
same as the weak eigenstates. The mixing between (down type) quark mass eigenstates was
described by the CKM  matrix \cite{ckm}. The mixing is expressed
in terms of a $3\times 3$ unitary matrix $V_{CKM}$ operating on
the down type quark mass eigenstates (d,s, and b):
\begin{equation}
\left(\begin{array}{ccc} d'\\s'\\b' \end{array}
\right)= \left(
\begin{array}{ccc}
V_{ud} & V_{us} & V_{ub}\\
V_{cd} & V_{cs} & V_{cb} \\
V_{td} & V_{ts} & V_{tb}
\end{array}
\right)
\left( \begin{array}{ccc} d\\s\\b \end{array} \right). \label{vckm}
\end{equation}

As a $3\times 3$ unitary matrix, the CKM mixing matrix $V_{CKM}$
is fixed by four parameters, one of which is an irreducible complex phase.
Using the generalized Wolfenstein
parametrization\cite{buras01}, $V_{CKM}$ takes the form
\begin{equation}
V_{CKM}= \left(           \begin{array}{ccc}
          1-\frac{\lambda^2}{2} & \lambda & A \lambda^3 (\bar{\rho}-i \bar{\eta})\\
          -\lambda & 1-\frac{\lambda^2}{2} & A \lambda^2 \\
          A \lambda^3 (1-\rhob-i \etab )&-A \lambda^2 & 1
          \end{array} \right) . \label{eq:vckm}
\end{equation}
where $A$, $\lambda$, $\rhob$ and $\etab$ are Wolfenstein parameters.

\begin{figure}[hbt]
\vspace{40pt}
\begin{minipage}[]{0.8\textwidth}
\centerline{\epsfxsize=0.8\textwidth \epsffile{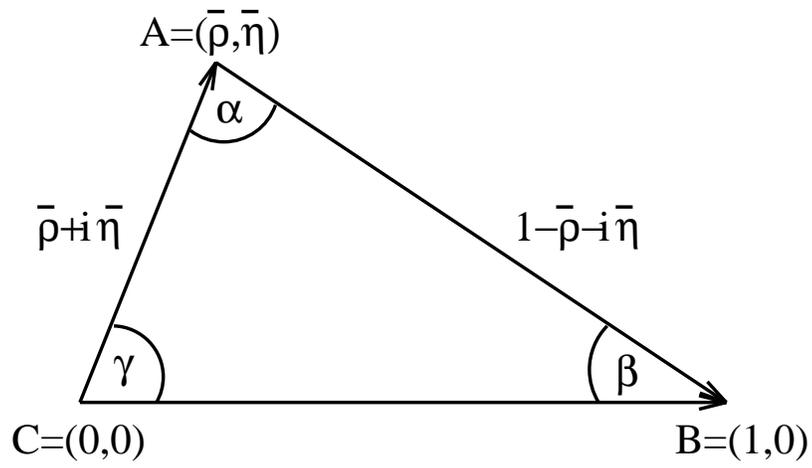}}
\vspace{20pt} \caption{Unitarity triangle in $\bar \rho , \bar
\eta$ plane.} \label{fig:fig1}
\end{minipage}
\end{figure}

 The unitarity of the CKM matrix implies  six ``unitarity
 triangle''. One of them applied to the first and third columns of
 the CKM matrix yields
\begin{equation}
V_{ud}V_{ub}^* + V_{cd}V_{cb}^* + V_{td}V_{tb}^* =0.
\end{equation}
This  unitary triangle is just a geometrical presentation of this equation
in the complex plane. We show it in the $\rhob-\etab$
plane, as illustrated in Fig.\ref{fig:fig1}.

The three unitarity angles are defined as
 \beq \alpha &=&
 \arg\left(-\frac{V_{tb}^*V_{td}}{V_{ub}^*V_{ud}}\right), \label{eq:alpha}\\
 \beta &=&
    \arg\left(-\frac{V_{cb}^*V_{cd}}{V_{tb}^*V_{td}}\right),
  \label{beta} \\
   \gamma &=& \arg\left(-\frac{V_{ub}^*V_{ud}}{V_{cb}^*V_{cd}}\right).
   \label{gamma} \eeq
The above definitions are independent of any kind of  parametrization of the CKM matrix
elements. Thus it is universal. In the Wolfenstein  parametrization,
 in terms of $(\rhob, \etab)$, $\sin(2\phi_i)$ ($\phi_i=\alpha,
 \beta, \gamma$) can be written as
 \beq
 \sin(2\alpha)&=&\frac{2\etab(\etab^2+\rhob^2-\rhob)}{(\rhob^2+\etab^2)((1-\rhob)^2
   +\etab^2)},\label{eq:sin2a}\\
 \sin(2\beta)&=&\frac{2\etab(1-\rhob)}{(1-\bar\varrho)^2 +
 \etab^2},  \label{eq:sin2b}\\
 \sin(2\gamma)&=&\frac{2\rhob \etab}{\rhob^2+\etab^2}.
 \eeq

The SM  predicts  the CP-violating asymmetries in the
time-dependent rates for initial $B^0$ and $\bar{B}^0$ decays to a
common CP eigenstate $f_{CP}$. In the case of $f_{CP} = \pi^+
\pi^-$, the time-dependent rate is given by
 \beq
 f_{\pi\pi}(\Delta
t) = \frac{e^{-|\Delta t|/ \tau_{B^0}}}{4 \tau_{B^0}} \left \{ 1+
q\cdot \left [ \spp \sin{(\Delta m_d \Delta t)} + A_{\pi\pi}
\cos{(\Delta m_d \Delta t)} \right ] \right \}  ,
 \eeq
 where
$\tau_{B^0}$ is the $B_d^0$ lifetime, $\Delta m_d$ is the mass
difference between the two $B_d^0$ mass eigenstates, $\Delta t
=t_{CP} -t_{tag}$ is the time difference between the tagged-$B^0$
($\bar{B}^0$) and the accompanying $\bar{B}^0$ ( $B^0$) with
opposite $b-$flavor decaying to $\pi^+ \pi^-$ at the time
$t_{CP}$, $q=+1$ ($-1$) when the tagging $B$ meson is a $B^0$
($\bar{B}^0$).
The CP-violating asymmetries $\spp$ and $A_{\pi\pi}$ have been defined as
\beq
S_{\pi\pi} = \frac{2 \im(\lpp)}{1+
|\lpp|^2}, \ \
A_{\pi\pi} &=& \frac{|\lpp|^2 - 1}{1 + |\lpp|^2},
\eeq
where the parameter $\lpp$ is
\beq
\lpp &=& \frac{V_{tb}^* V_{td}}{V_{tb}V_{td}^*} \left [ \frac{V_{ub}
V_{ud}^* \tpp e^{i\delta_1} - V_{tb} V_{td}^* \ppp e^{i\delta_2}
}{ V_{ub}^* V_{ud} \tpp e^{i\delta_1} - V_{tb}^* V_{td} P\ppp
e^{i\delta_2}}\right ] \non &=& e^{2 i \alpha} \left [ \frac{1+ r
e^{i(\delta - \alpha)}}{ 1+ r e^{i(\delta + \alpha)}} \right ]  ,
\eeq
with
\beq r = \left | \frac{P_{\pi\pi}}{T_{\pi\pi}} \right |
\left | \frac{V_{tb} V_{td}^*}{V_{ub}V_{ud}^*}\right |, \ \ \delta
= \delta_2 - \delta_1, \label{eq:rth} \eeq
where $\tpp$ and $\ppp$ describe the ``Tree" and  penguin contributions to the $\bpp$
decay, and $\delta$ is the difference of the corresponding strong phases of tree and
penguin amplitudes.

By explicit calculations, we find that
\beq
\spp &=& \frac{ \sin{2\alpha} + 2 r \cos{\delta } \sin{\alpha} }{
1 + r^2 + 2 r \cos{\delta} \cos{\alpha}  }, \label{eq:sppth}\\
A_{\pi\pi} &=& \frac{ 2 r \sin{\delta} \sin{\alpha}}{ 1 + r^2 + 2
r \cos{\delta} \cos{\alpha} } .\label{eq:cppth}
\eeq
If we neglect the penguin-diagram (which is expected to be smaller than the tree
diagram contribution), we have $A_{\pi \pi}=0$, $S_{\pi\pi}=\sin(2\alpha)$. That
means we can measure the $\sin (2\alpha)$ directly from $\bpp$ decay. This is the reason why
$\bpp$ decay was assumed to be the best channel to measure CKM
angle $\alpha$ previously. With penguin contributions, we have
$A_{\pi \pi}\neq 0, S_{\pi\pi}=\sin(2\alpha_{eff})$, where
$\alpha_{eff}$ depends on the magnitude and strong phases of the
tree and  penguin amplitudes. In this case, the CP asymmetries can not tell the size
of angle $\alpha$ directly. A method has been proposed to extract CKM angle $\alpha$
using $B^+\to \pi^+\pi^0$ and $B^0\to \pi^0\pi^0$ decays together with $\bpp$
decay  by the isospin relation \cite{gronau2}. However, it will take quite some
time for the experiments  to measure the three channels together.

\section{Constraint on $\alpha$ and $\delta$}\label{sec-3}

In this section, we will show that the only measured CP asymmetries of
$\bpp$ decay can at least provide some constraint on the angle $\alpha$.

 From Eq.(\ref{eq:sppth},\ref{eq:cppth}), one can see that the
asymmetries $\spp$ and $A_{\pi\pi}$ generally depend on three
``free" parameters: the CKM angle $\alpha$ with $\alpha=[0, \pi]$,
the strong phase $\delta$ with $\delta=[-\pi,\pi]$ and the ratio
$r$ as defined in Eq.(\ref{eq:rth}). We can not solve out these
two equations with three unknown  variables. However, by the
following study, we can at least give some constraint on the angle
$\alpha$ and strong phase $\delta$. Since the penguin
contributions are loop order corrections ($\alpha_s$ suppressed)
comparing with the tree contribution, we can assume $0 < r <
0.5$, in a reasonable range.

Now we are ready to extract out $\alpha$ through the general parameterization of
$\spp$ and $A_{\pi\pi}$ in terms of $(\alpha, \delta, r)$ as given in
Eqs.(\ref{eq:sppth},\ref{eq:cppth}). As discussed previously \cite{rosner}, there may exist
some discrete ambiguities between $\delta$ and $\pi - \delta $ for the
mapping of $\spp$ and $\app$ onto the $\delta-\alpha$ plane.

At first, because of the positiveness of $A_{\pi\pi}^{exp}$ at $1\sigma$ level and the fact that
$\sin{\alpha} >0$ for $\alpha=(0,\pi)$,
the range of $-\pi <  \delta  <  0 $ and $\delta=0, \pm \pi$ are excluded.
and therefore only the range of $0^\circ <  \delta < 180^\circ$ need to be considered
here.

For the special case of $\delta=90^\circ$, the discrete ambiguity between $\delta$ and
$\pi-\delta$ disappear and the expressions of $\spp$ and $\app$ can be
rewritten as
\beq
\spp &=& \frac{ \sin{2\alpha} }{ 1 + r^2 },\label{eq:spp90} \\
\app &=& \frac{ 2 r \sin{\alpha}}{ 1 + r^2 } .\label{eq:app90}
\eeq
The range of $0^\circ \leq \alpha \leq 90^\circ $ is excluded by the negativeness of $\spp^{exp}$,
and the range of $\sin{\alpha} < 0.19(1+r^2)/r $ and  $\sin{\alpha} > 0.38(1+r^2)/r $ are
excluded by the measured  $\app = 0.57 \pm 0.19$ at the $1\sigma$ level.

\begin{figure}[t]
\vspace{-40pt}
\begin{minipage}[]{\textwidth}
\centerline{\epsfxsize=\textwidth \epsffile{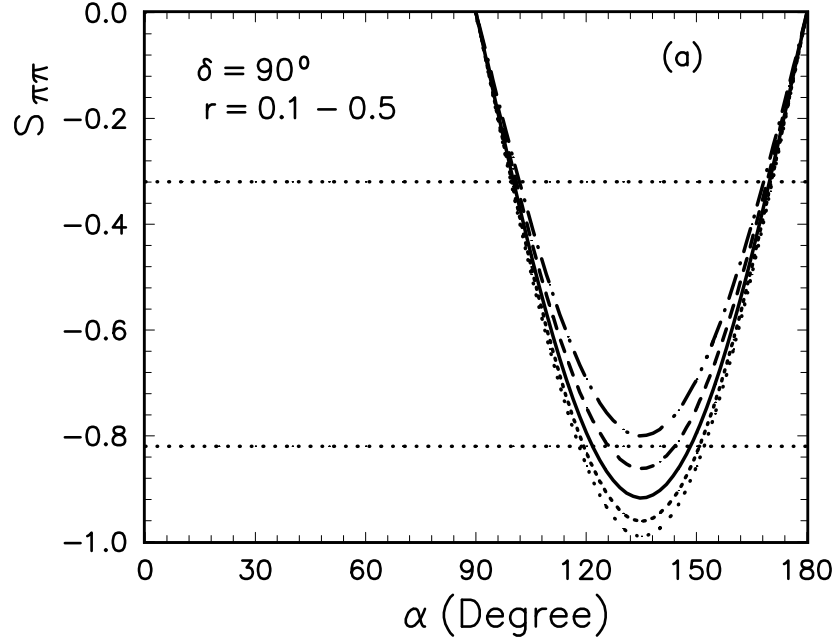}}
\vspace{-60pt}
\centerline{\epsfxsize=\textwidth \epsffile{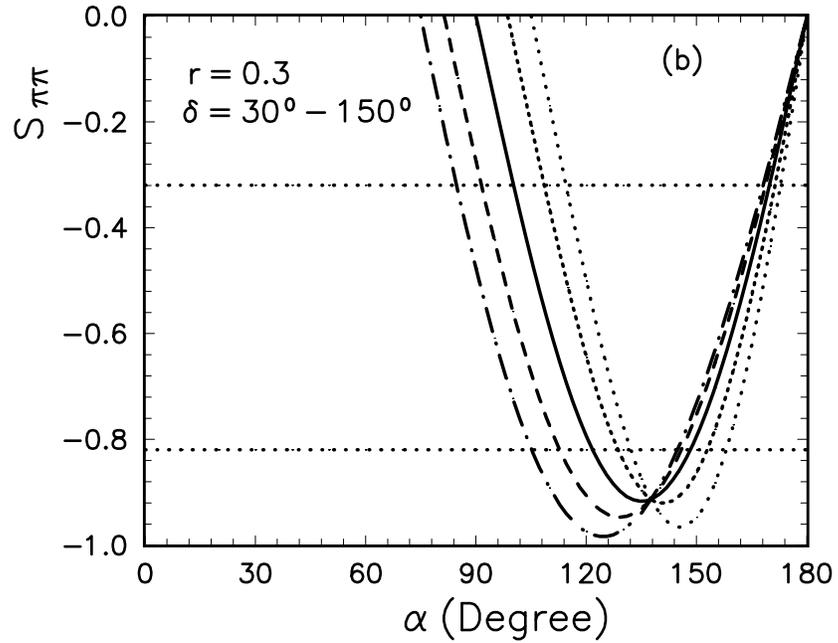}}
\vspace{-20pt}
\caption{Plots of $\spp$ vs the angle $\alpha$. (a) The dots-, tiny dashed-, solid-,
dashed and
dash-dotted curves correspond to $r=0.1$, $r=0.2$, $0.3$ $r=0.4$ and $0.5$, respectively.
(b) The five curves from left to right  are for $\delta=150^\circ, 120^\circ$,
$90^\circ$, $60^\circ$ and $30^\circ$ respectively.
The band between two horizontal dotted lines shows the experimental $1\sigma$ allowed range
 $ -0.82 \leq \spp^{exp} \leq -0.32$. }
\label{fig:fig2}
\end{minipage}
\end{figure}

In Fig.\ref{fig:fig2}a, we show the  $\alpha$ dependence of $\spp$ for given $\delta=90^\circ$
and  for $r=0.1$ (dotted curve), $0.2$ (tiny-dashed curve),  $0.3$ (solid curve),
$0.4$ (dashed curve) and $0.5$ (dash-dotted curve), respectively.
The band between the two horizontal dots lines shows the allowed range from the measured
$\spp^{exp} = -0.57 \pm 0.25$ at $1\sigma$ level.
Fig.\ref{fig:fig2}b shows the $\alpha$ dependence of $\spp$ for fixed
$r=0.3$ and for $\delta=30^\circ$ (dotted curve),$60^\circ$ (tiny-dashed curve),
$90^\circ$ (solid curve), $120^\circ$ (dashed curve), and $150^\circ$ (dash-dotted curve),
respectively. The differences between the curves of $\delta=30^\circ$ and $\delta
=150^\circ$ ( $\delta=60^\circ$ and $\delta =120^\circ$) show  the effects
of discrete ambiguity between $\delta$ and $\pi-\delta$.

The constraint on the CKM angle $\alpha$ from the measured $\spp$ alone can be read
off directly from figure 2. For $r=0.3$, for example, the allowed ranges for the CKM angle
$\alpha$ are
\beq
109^\circ \leq \alpha \leq 128^\circ \bigvee 153^\circ \leq \alpha \leq 171^\circ
\label{eq:limit1}
\eeq
for $\delta=60^\circ$, and
\beq
101^\circ \leq \alpha \leq 121^\circ \bigvee 149^\circ \leq \alpha \leq 169^\circ
\label{eq:limit2}
\eeq
for $\delta=90^\circ$, and
\beq
92^\circ \leq \alpha \leq 112^\circ \bigvee 146^\circ \leq \alpha \leq 168^\circ
\label{eq:limit3}
\eeq
for $\delta=120^\circ$. In general, the current experimental measurements of
$\spp$ prefer to $\alpha > 90^\circ$.

\begin{figure}[t]
\vspace{-40pt}
\begin{minipage}[]{\textwidth}
\centerline{\epsfxsize=\textwidth \epsffile{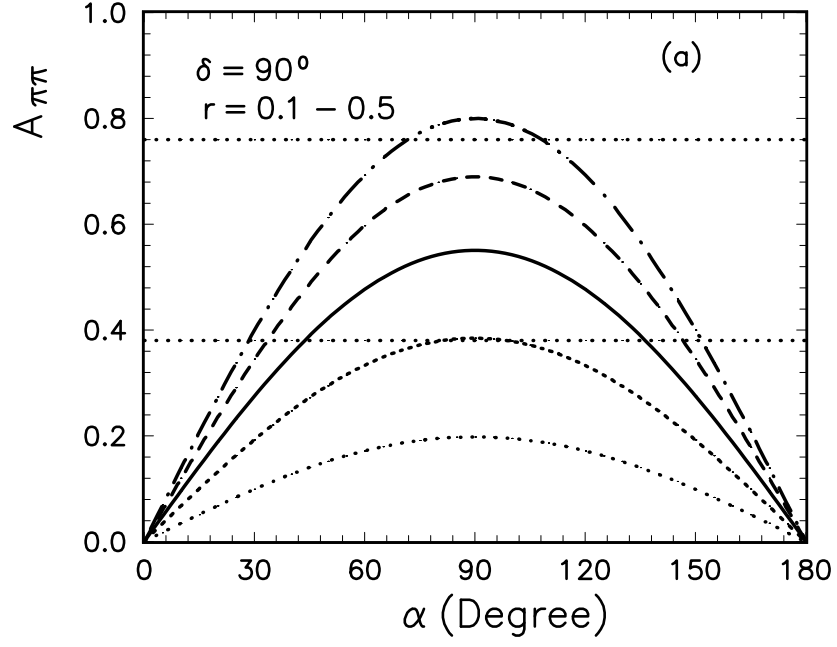}}
\vspace{-60pt}
\centerline{\epsfxsize=\textwidth \epsffile{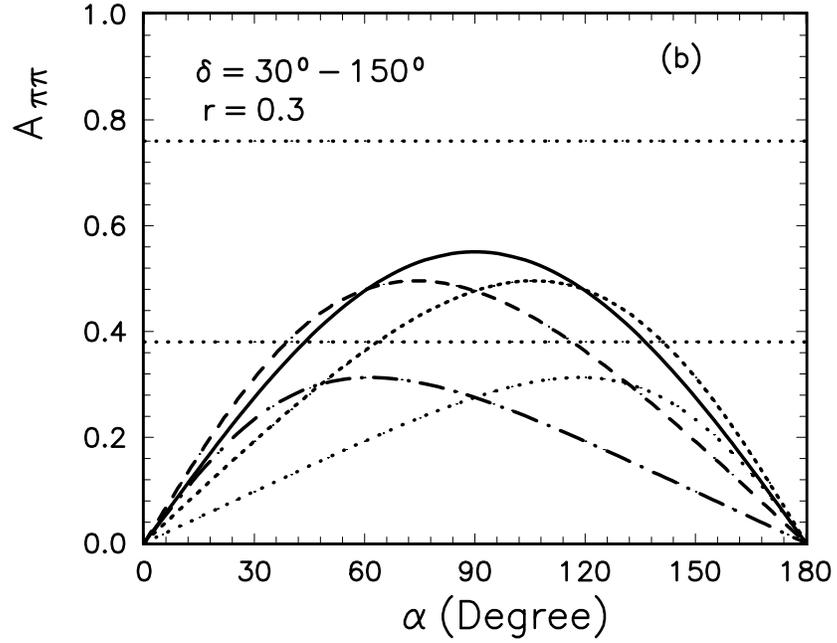}}
\vspace{-20pt}
\caption{Plots of $A_{\pi\pi}$ vs the angle $\alpha$.  In (a) the dots-, tiny dashed-, solid-,
dashed and dash-dotted curves correspond to $r=0.1$, $0.2$, $0.3$ $0.4$ and $0.5$,
respectively. In (b) the same ordered curves are for $\delta=30^\circ$, $\delta=60^\circ$,
$\delta=90^\circ$, $130^\circ$ and $150^\circ$ respectively.
The band between two horizontal dotted lines shows the experimental $1\sigma$ allowed range
 $0.38 \leq A_{\pi\pi}^{exp} \leq 0.76$. }
\label{fig:fig3}
\end{minipage}
\end{figure}

In Fig.\ref{fig:fig3}a, we show the  $\alpha$ dependence of $A_{\pi\pi}$ for given
$\delta=90^\circ$ and  $r=0.1$ (dotted curve), $0.2$ (tiny-dashed curve),  $0.3$ (solid curve),
$0.4$ (dashed curve) and $0.5$ (dash-dotted curve), respectively.
The band between two horizontal dots line shows the allowed region by the measured
$\app^{exp} = 0.57 \pm 0.19$ at $1\sigma$ level.
Fig.\ref{fig:fig3}b shows the $\alpha$ dependence of $A_{\pi\pi}$ for given
$r=0.3$ and for  $\delta=30^\circ$ (dotted curve),$60^\circ$ (tiny-dashed curve),
$90^\circ$ (solid curve), $120^\circ$ (dashed curve), and $150^\circ$
(dash-dotted curve), respectively.
It is easy to see that most parts of the allowed ranges of $\alpha$ as given in
Eqs.(\ref{eq:limit1}-\ref{eq:limit3}) can be excluded by the inclusion of
measured $\app$. The second solutions as given in Eqs.(\ref{eq:limit1}-\ref{eq:limit3}) will
be  removed by taking the measured $\app$ into account.
For the case of $r=0.3 $ and $\delta \leq 30^\circ$ or $\delta \geq 150^\circ$, the whole
range of $\alpha$ will be excluded by the measured $\spp$ and $\app$, as illustrated
in Fig.\ref{fig:fig3}b.

From the above analysis, we can see that the strong constraint  on
CKM angle $\alpha$ can be obtained by using the experimental
measurements of $\spp$ and $\app$ as well as the ratio $r$. With
the rapid increase of the $B \bar{B}$ pair production and decay
events collected at B-factory experiments, the difference between
the central value of $\spp$ and $\app$ and the experimental
uncertainties will become smaller within two years. For the third
input parameter $r$, it can be fixed through available data or
reliable theoretical considerations.

From  Eqs.(\ref{eq:ssb-av2}) and (\ref{eq:sin2b}), the measured $\ssb$
leads to an equation between $\rhob$ and $\etab$,
 \beq
 \etab = (1-\rhob)
 \xi = (1-\rhob) \frac{1 \pm \sqrt{1- \sin^2(2\beta)}}{\ssb}.
 \label{eq:etab3}
 \eeq
 The solution with the $``+"$ sign in the numerator of $\xi$ is  totally inconsistent with the
 global fit results and can be dropped out. Numerically,
 \beq
 \xi= \left
 (1 -\sqrt{1- \sin^2(2\beta)} \right )/\ssb =
 0.48^{+0.09}_{-0.07},
 \label{eq:xi}
 \eeq
 for the measured $\ssb=0.78 \pm 0.08$. There exist quite a lot of information about the
 CKM matrix elements as reported by the Particle Data Group \cite{pdg00} and
other recent papers \cite{hocker01,he01,falk02,xiao02,atwood01}.
 The  parameter $\lambda=|V_{us}|$ is known from $K_{l3}$ decay with
 good precision
 \beq
 \lambda= 0.2196 \pm 0.0023. \label{eq:lambda}
 \eeq

 In terms of $(\rhob, \etab)$, the parameter
 $r$ as defined in Eq.(\ref{eq:rth}) can be rewritten as
 \beq
 r &=& z \frac{\sqrt{(1-\rhob)^2 + \etab^2}}{\sqrt{ \rhob^2 + \etab^2 }
 \left (  1-\frac{\lambda^2}{2}\right ) } =
 \frac{z}{1-\frac{\lambda^2}{2}} \frac{(1-\rhob) \sqrt{ 1+ \xi^2}
 }{ \sqrt{\rhob^2 + (1-\rhob)^2  \xi^2}} ,\label{eq:rth2}
 \eeq
 where
 $z=|\ppp/\tpp|$ measures the relative size of tree and penguin  contribution to the
 studied decay. From general considerations, $z$ may be around $20\%$. By employing
 the QCD factorization  approach \cite{bbns3} and/or the perturbative QCD approach \cite{pqcd},
one can fix $r$ to a rather good degree.
By using the QCD factorization approach, the estimated  value of  $|\ppp/\tpp|$ is
found \cite{bbns3} to be
\beq
  z=0.285 \pm 0.077, \label{eq:z}
\eeq
 where the contribution from the weak annihilation has been
 taken into account and the dominate error comes from the
 uncertainties of $m_s$ and the renormalization scale $\mu$  \cite{bbns3}.
 From the numbers as given in Eqs.(\ref{eq:xi},\ref{eq:lambda},\ref{eq:z})
 and $\rhob = 0.20\pm  0.16$, we have numerically
 \beq
 r = 0.31 \pm 0.09 (\Delta z) \pm 0.01 (\Delta \xi) ^{+0.01}_{-0.03} (\Delta \rhob)
 \pm 0.0002 (\Delta \lambda) = 0.31 \pm 0.10, \label{eq:rth3}
 \eeq
 Here the estimated result $r \leq 0.41$ is in good agreement with our
 general argument of $ r< 0.5$. Thus our analysis in this paper is
 meaningful.

\begin{figure}[t]
\vspace{-40pt}
\begin{minipage}[]{\textwidth}
\centerline{\epsfxsize=\textwidth \epsffile{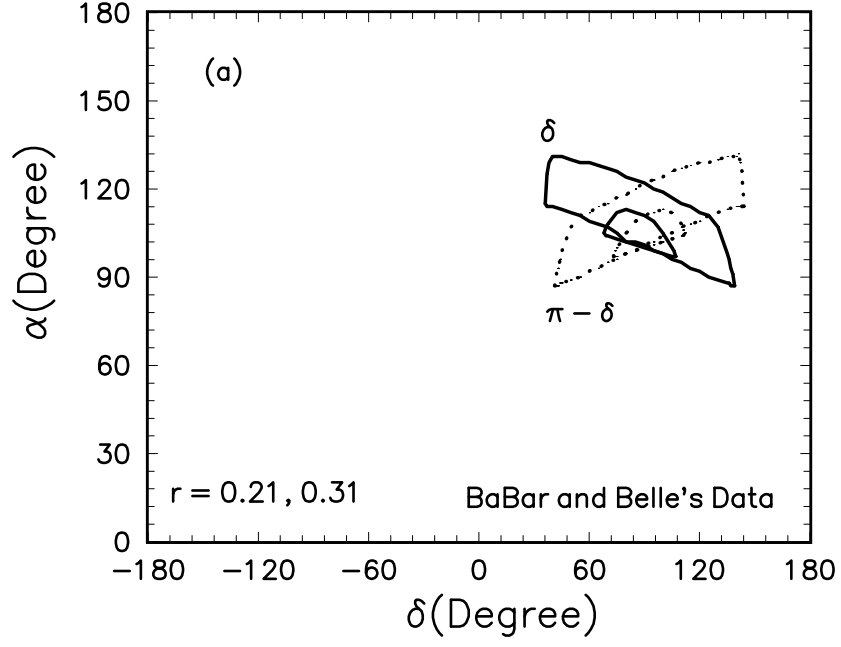}}
\vspace{-60pt}
\centerline{\epsfxsize=\textwidth \epsffile{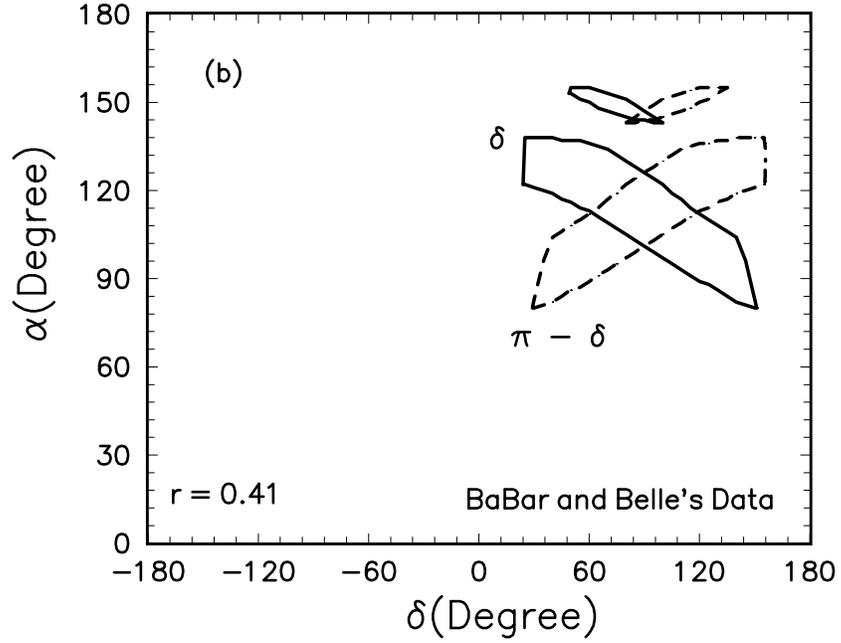}}
\vspace{-20pt}
\caption{Contour plots of the
asymmetries $\spp$ and $\app$ versus the strong phase $\delta$ and
CKM angle $\alpha$ for $r=0.21$ (the small circles in (a)) and $0.31$ (the large
circles in (a)), and $0.41$ (b), respectively. The dotted circles show the
effects of discrete ambiguity. The regions inside each circle are still allowed by
both $ -0.82 \leq \spp^{exp} \leq -0.32$ and $0.38 \leq
\app^{exp} \leq 0.76$, which is the experimental $1\sigma$
allowed range. } \label{fig:fig4}
\end{minipage}
\end{figure}

The common range of $\alpha$ allowed by both the measured $\spp$ and $\app$ is what
we try to find. Fig.\ref{fig:fig4} shows the contour plots of the CP asymmetries $\spp$ and
$\app$ versus the strong phase $\delta$ and CKM angle $\alpha$ for $r=0.21$
(the small circles in (a)), $0.31$ (the larger circles in (a)) and $0.41$ (circles in (b)),
respectively. The regions inside each circle
are still allowed by both $\spp^{exp} = -0.57 \pm 0.25$ and $\app^{exp} = 0.57 \pm 0.19$
(experimental $1\sigma$ allowed ranges). The discrete ambiguity between
$\delta$ and $\pi -\delta$ are shown by the solid and dotted circles in
Fig.\ref{fig:fig4}. For $\delta =90^\circ$, such discrete ambiguity disappear.

If we take the theoretically fixed value of $r=0.31 \pm 0.10$ as the reliable estimation of $r$,
the constraint on the CKM angle $\alpha$ and the strong phase $\delta$
can be read off directly from
Fig.\ref{fig:fig4}. Numerically, the allowed regions for the CKM angle $\alpha$ and
the strong phase $\delta$ are
\beq
97^\circ \leq \alpha \leq 113^\circ, \ \  68^\circ \leq \delta \leq 112^\circ
\eeq
for $r=0.21$,  and
\beq
87^\circ \leq \alpha \leq 131^\circ,  \ \  36^\circ \leq \delta \leq 144^\circ
\eeq
for $r =0.31$,  and finally
\beq
&&80^\circ \leq \alpha \leq 138^\circ \bigvee 143^\circ \leq \alpha \leq 155^\circ, \non
&&24^\circ \leq \delta \leq 151^\circ
\label{eq:limit41}
\eeq
for $r =0.41$. There is a twofold ambiguity for the determination
of angle $\alpha$ for $r\approx 0.4$.  In fact, the CKM angle $\alpha$ in the
second region in Eq.(\ref{eq:limit41}) is  too big to be consistent with the
standard model unitarity relation: $\alpha + \beta + \gamma =180^\circ$.

One can see from Fig.\ref{fig:fig4} that if we take the weighted-average of the BaBar and
Belle first measurements of the asymmetries $\spp$ and $\app $ as the reliable measured
values of $\spp$ and $\app$, we can obtain strong constraint on both the strong phase $\delta$
and the CKM angle $\alpha$. Even we consider the  uncertainties of
input parameters, most part of the parameter space is also excluded.

\begin{figure}[t]
\vspace{-40pt}
\begin{minipage}[]{\textwidth}
\centerline{\epsfxsize=\textwidth \epsffile{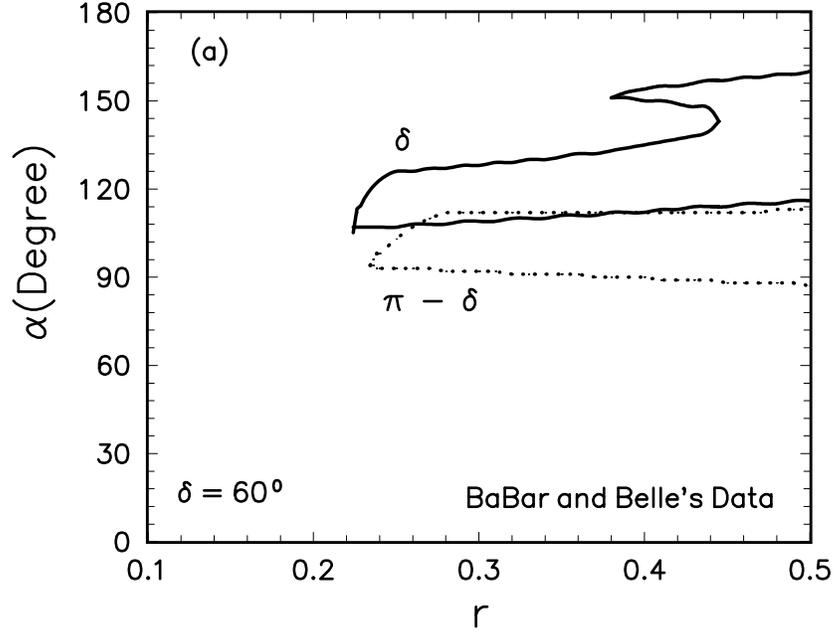}}
\vspace{-60pt}
\centerline{\epsfxsize=\textwidth \epsffile{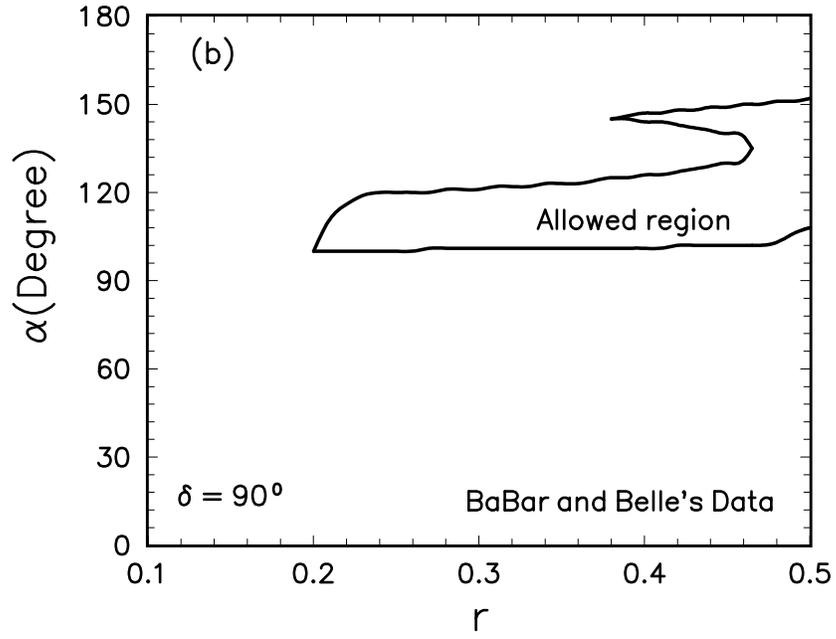}}
\vspace{-20pt}
\caption{Contour plots of the asymmetries $\spp$ and $\app$ versus the CKM angle $\alpha$
and the ratio $r$ for $\delta=60^\circ$ and $120^\circ$ (Fig.5a) and $90^\circ$ (Fig.5b),
respectively. The dotted semi-closed curve in (a) shows the effects of discrete ambiguity.
The regions inside the semi-closed curves are still allowed by the data.}
\label{fig:fig5}
\end{minipage}
\end{figure}

In order to show more details of the $r$ dependence of the constraint on
$\alpha$, we draw Fig.\ref{fig:fig5}. The semi-closed regions as shown in Fig.5a
(for $\delta =60^\circ$ and $120^\circ$)
and Fig.5b (for $\delta=90^\circ$) are still allowed by the measured $\spp$ and $\app$
as given in  Eq.(\ref{eq:data}).  As shown in Fig.\ref{fig:fig5}, the region of $r\leq 0.2$
is excluded by the data. The effects of discrete ambiguity are also
shown in Fig.5. The solid semi-closed region in Fig.5a corresponds to $\delta=60^\circ$, while
the dotted semi-closed region refers to $\pi-\delta=120^\circ$. For
$\delta=90^\circ$, such discrete ambiguity disappears.

 As discussed in previous section, there are some discrepancy between the BABAR and Belle
 measurements of $\spp$ and $\app$ (or $\cpp$ ). If we use  Belle's measurement of
 $\spp$ and $\app$ only, and  take the direct sum of statistic and systematic errors as
 the total $1\sigma$ error, then the experimental limits on both $\spp$ and $\app$
 take the form
 \beq
\spp^{exp} \leq -0.67, \ \ \app^{exp} \geq 0.54.\label{eq:datab}
 \eeq
 The corresponding contour plots of the asymmetries $\spp$ and $\app$ versus the
 strong phase $\delta$ and CKM angle $\alpha$ are illustrated in Fig.\ref{fig:fig6}
 for $r=0.36$ (the small solid circle), $0.41$ (the middle-sized solid circle)
 and $0.51$ (the large solid circle), respectively.
 For $r\leq 0.32$, the whole $``\delta-\alpha"$ plane is excluded. The dotted
 circles correspond to the discrete ambiguity between $\delta$ and
 $\pi=\delta$.
 Numerically, we find that the allowed ranges for the CKM angle $\alpha$ and the strong
 phase $\delta$ are
 \beq
 108^\circ \leq \alpha \leq  130^\circ, \ \  54^\circ \leq \delta \leq  126^\circ
 \eeq
for $r=0.36$, and
 \beq
 104^\circ \leq \alpha \leq  139^\circ,  \ \  42^\circ \leq \delta \leq  138^\circ
 \eeq
for $r=0.41$, and finally
 \beq
 95^\circ \leq \alpha \leq  152^\circ,  \ \  28^\circ \leq \delta \leq  152^\circ
 \eeq
for $r=0.51$, although we do not expect so large value of the  ratio $r$.

If we use the Belle's measurement of $\spp$ and $\app$ only, and
 take the square root of the statistic  and systematic errors as the total $1\sigma$
 error, then the experimental limits on both $\spp$ and $\app$ will be
 \beq
 \spp^{exp} \leq -0.80, \ \ A_{\pi\pi}^{exp} \geq 0.62 . \label{eq:datac}
 \eeq
 The whole $``\delta-\alpha"$ plane will be excluded even for $r=0.51$.
 In other word, the Belle result has to be changed in the future,
 otherwise, new physics may be required to explain the data.

Since the discrete ambiguity between $\delta$ and $\pi -\delta$ vanishes when $\delta =\pi/2$,
the contour plots as shown in Figs.4 and 6 are symmetric with respect to the axis of
$\delta=90^\circ$ in the $\delta-\alpha$ plane.
Such discrete ambiguity can alter the constraints on $\delta$ by about $7^\circ$, but has
little effect on the possible limits on the CKM angle $\alpha$ derived from the measured
$\spp$ and $\app$ if we fix the value of $r$ and treat $\delta$ as a free parameter
varying in the range of $0^\circ < \delta < 180^\circ$ as can be seen in Figs. 4 and 6.

It is worth to mention that the constraint on the angle $\alpha$ from one recent global fit
is $82^\circ \leq \alpha \leq 126^\circ$ as given in Ref.\cite{hocker01}.
The constraint from measured $\spp$ and $\app$ is comparable or stronger than the global fit
result.

\section{Conclusion}

In this paper, we studied the $\bpp$ decay and try to find constraint
on the CKM angle $\alpha$ and the strong phase $\delta$ from the measured asymmetries $\spp$ and
$\app$ as reported by the BaBar and Belle Collaborations.

\begin{figure}[tb]
\vspace{-40pt}
\begin{minipage}[]{\textwidth}
\centerline{\epsfxsize=\textwidth \epsffile{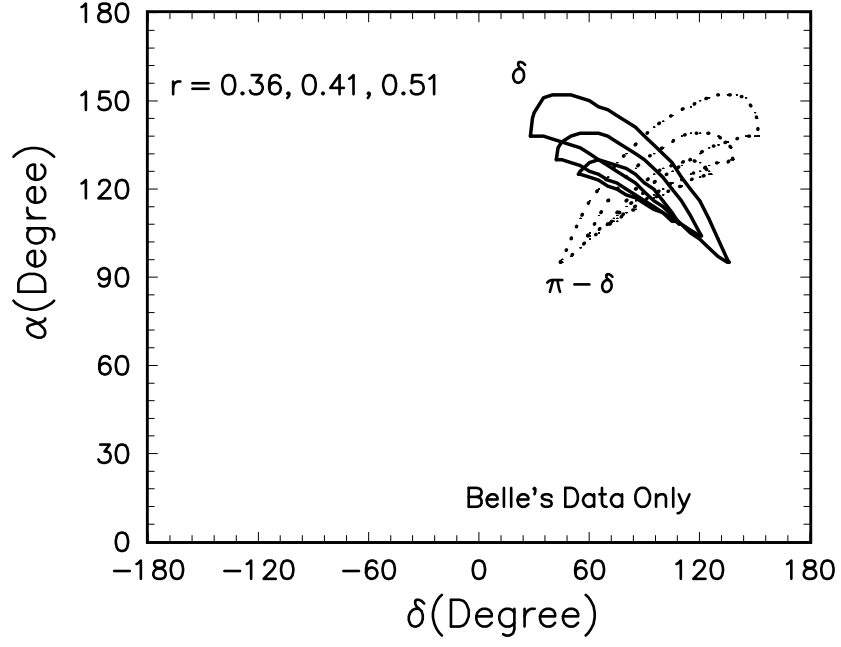}}
\vspace{-20pt}
\caption{Contour plot of the asymmetries $\spp$ and $A_{\pi\pi}$ versus the strong phase $\delta$
and CKM angle $\alpha$ for $r=0.36$ (small solid circle), $0.41$ (middle-sized solid circle )
and $0.51$ (large solid circle ), respectively.
The dotted circles show the effects of discrete ambiguity.
The regions inside each circle is still allowed by the Belle's
limits $\spp^{exp} \leq -0.67$ and $\app^{exp} \geq 0.54$. }
\label{fig:fig6}
\end{minipage}
\end{figure}

If we take the weighted-average
of BABAR and Belle's measurements of $\spp$ and $\app$ as the measured results,
strong constraint on both the CKM angle $\alpha$ and the strong phase $\delta$ can be obtained.
The range of $\delta \leq 0 $ is excluded by the positiveness of measured $\app$.
The range of $0^\circ \leq \alpha \leq 90^\circ$ is excluded by the negativeness
of measured $\app$ for $\delta = 90^\circ$. Within the parameter space of
$\spp=-0.57 \pm 0.25$, $\app = 0.57 \pm 0.19$ and $r =0.31 \pm 0.10$, most part of the
$``\delta-\alpha"$ plane is excluded, as shown in Figs.(\ref{fig:fig3}-\ref{fig:fig6}).
For fixed $r =0.31$, for example, the ranges of $87^\circ \leq \alpha \leq 131^\circ$ and
$36^\circ \leq \delta \leq 144^\circ$ are allowed by $1\sigma$ of the averaged
$\spp$ and $\app$. In general the data prefer $ \alpha > 90^\circ$.

The discrete ambiguity between $\delta$ and $\pi-\delta$ will disappear for
$\delta=90^\circ$ and has little effects on the possible limits on the CKM
angle $\alpha$ if we fix the value of $r$ and treat $\delta$ as a free
parameter varying in the range of $0^\circ < \delta < 180^\circ$, as shown in
Figs. 4 and 6.

If we consider only  Belle's measurements, a very narrow range in the $``\delta-\alpha"$ plane is
allowed, as illustrated in Fig.\ref{fig:fig6}.
The limits on $\alpha$ and $\delta$ are
$ 104^\circ \leq \alpha \leq  139^\circ$ and $42^\circ \leq \delta \leq
138^\circ$ for $ 0.32 \leq r \leq 0.41$;
Considering the previous $\sin 2\beta $ measurement
$\beta =26^\circ \pm 4^\circ$), we can conclude that the other CKM angle
$\gamma$ should be smaller than $90^\circ$.

We know that the current  data of $\spp$ and $\app$ are still some kind of
preliminary experimental measurements with large uncertainties.
The apparent large difference between the BABAR and Belle measurements and the corresponding
experimental uncertainties will become smaller along with the rapid increase of
the observed $B$ decay events.  Therefore we are able to extract out the angle $\alpha$
with a good accuracy soon.

\vspace{0.5cm}

\section*{ACKNOWLEDGMENTS}
This work is partly supported by National Science Foundation of
China under Grant No.~90103013 and 10135060.
Z.J.~Xiao acknowledges the support by the National Natural Science Foundation of
China under Grant No.~10075013, and by the Research Foundation of Nanjing Normal
University under Grant No.~214080A916.


\newpage

\newpage

\newpage

\newpage

\newpage

\newpage

\end{document}